\documentclass[
    ,final            
  ]
  {aipproc}
\layoutstyle{6x9}
\usepackage[mathcal]{eucal}
\usepackage{amssymb}
\usepackage{amsmath,epsfig}
\usepackage{bbm}
\font\zul=cmbx6 at 6pt

\begin{document}

\title
      [Quantum $\kappa$-deformations of D=4 relativistic supersymmetries]
      {Quantum $\kappa$-deformations of D=4 relativistic supersymmetries\thanks{Supported by KBN grant
       1P03B 01828}}


\author{J. Lukierski}{
address={Institute of Theoretical Physics \\
50-205 Wroc\l aw,
pl. Maxa Borna 9, Poland}, email={lukier@ift.uni.wroc.pl}    }

\copyrightyear  {2006}

\begin{abstract}
We describe the  quantum $\kappa$-deformation of
super-Poincar\'{e} algebra, with fundamental mass-like deformation
parameter $\kappa$.
We shall describe the result in graded bicrossproduct basis,
with classical Lorentz superalgebra sector which includes
half of the supercharges.
\end{abstract}

\date{\today}
\maketitle

\subsection{Introduction}

In  order to define the quantum deformations one should consider
the deformed Lie algebra $A = U_\zeta(\widehat{g})$ as an algebraic
sector of the Hopf algebra $H=(A,\Delta,S,\epsilon)$, where $A$
 denotes associative algebra, the coproduct $\Delta$
 ($\Delta : A \to A \otimes A$) describes  the coalgebra structure,
 $S$ defines the  coinverse (antipode) ($S: A \to A$) and the counit $\epsilon$ is a complex
  functional on $A$ ($\epsilon : A \to C$) (see e.g. \cite{maj}).
In classical Hopf-Lie algebra with Lie algebra basis $I_i
\in \widehat{g}$ where $[I_i, I_j]=c_{ij}^{\ \ k} I_k$ the generators
$I_i$ are endowed with the primitive (Abelian) coproducts
 $\Delta^{(0)} (I_i)$
\begin{equation}\label{eql2}
I_i \quad \to \quad \Delta^{(0)} (I_i) = I_i \otimes {\mathbbm 1}
+ {\mathbbm 1} \otimes I_i
\end{equation}
Following Drinfeld \cite{dri} the quantum symmetries in deformed
Lie-algebraic framework are described by noncocommutative Hopf
algebras, with nonsymmetric coproducts
\begin{equation}\label{eql3}
\tau (\Delta (I_i )) \neq \Delta(I_i)\, ,
\end{equation}
where $\tau$ denotes the flip operator
 ($\tau (a \otimes b) = b \otimes a$). If the coproduct is
symmetric i.e. $\tau(\Delta (I_r))= \Delta(I_r)$
  we deal with  the ``fake" quantum deformations, which can be
obtained by nonlinear transformation of the classical Lie algebra
generators: $I_i \to J_i = J_i (\overrightarrow{I})$
($\overrightarrow{I}=(I_1 \ldots I_N ; N= dim \widehat{g}$). The
coproduct for such nonlinearly realized classical symmetry
 in arbitrary basis
   remains symmetric.

In our note we shall discuss the  quantum
relativistic supersymmetries characterized by mass-like deformation
 parameter (so-called $\kappa$-deformations). The general  quantum deformations of
supersymmetries are described by the noncocommutative $Z_2$-graded Hopf superalgebras
 (see e.g. \cite{cha,kho}.
We recall that the postulates
 for the Hopf superalgebra $\widetilde{H}=(\widetilde{A},
\widetilde{\Delta},\widetilde{S},\widetilde{\epsilon})$ are obtained
from the ones for Hopf algebra $H=(A,\Delta, S, \epsilon)$ by introducing the sign factors
in accordance with the rules of supermathematics (see e.g. \cite{maj}).
In particular  one introduces $Z_2$-graded super-Lie bracket
$ [a,b] = ab = (-1)^{{\rm grad}\, a \ {\rm grad}\, b} ba $
 (${\rm grad} \, a = 0$ for bosonic (even) $a$,
 ${\rm grad} \, a = 1$ for fermionic (odd) $a$)
 and the multiplication rule of $Z_2$-graded tensor products
 $(a \otimes b) (c \otimes d) =   (-1)^{{\rm grad}\, b \cdot {\rm grad}\, c} (ac\otimes bd)$.
An important class of quantum supersymmetries is generated from
classical Hopf-Lie superalgebra
 ($\widetilde{H}^{(0)} = (\widetilde{A}^{(0)}, \widetilde{\Delta}^{(0)},
\widetilde{S}^{(0)},\widetilde{\epsilon}^{(0)}$) by Drinfeld twist
$\widetilde{F}$ which is an even element
  of superalgebra $\widetilde{A}^{(0)}
\otimes \widetilde{A}^{(0)} \
  (\widetilde{F}
= \sum \widetilde{F}_{(1)} \otimes \widetilde{F}_{(2)}$, where ${\rm grad }\widetilde{F}_{(1)}
+ {\rm grad} \widetilde{F}_{(2)} = 0 \  \hbox{or \ 2})$.
In twisted Hopf-Lie superalgebras we  modify only the classical
 formulae for coproducts and antipodes.

Below we shall present the $\kappa$-deformation of superPoincar\'{e} algebra  which is
 the superextension of standard $\kappa$-deformation of Poincar\'{e}
  algebra \cite{luk,maj2}.
  We shall describe this deformation in bicrossproduct basis with classical Lorentz
  algebra sector \cite{maj2}, what   is suitable for the supersymmetrization of recently
   studied Doubly Special
  Relativity framework (see e.g. \cite{kow}).

  \subsection{The example of quantum-deformed SUSY: superextension of standard
   $\kappa$-deformation of Poincar\'{e} algebra }

  Let us recall the $\kappa$-deformed Poincar\'{e} algebra \cite{luk} in bicrossproduct
   basis \cite{maj2}:
   \begin{equation}\label{eql15}
   {\mathcal P}_{4} = O(3,1)
    \ltimes T_{4}
\mathop{\longrightarrow}\limits_{ \kappa < \infty } \
   \ {\mathcal P}^{\kappa}_{4} = O(3,1)
\mathop{\triangleright\!\!\!\!\blacktriangleleft}\limits_{ \kappa } \
    T_4 ^{\ \kappa}
   \end{equation}
where $T^\kappa_{4}$  is a Hopf algebra of $\kappa$-deformed Abelian translation
 generators ($[P_\mu, P_\nu]=0$)
 \begin{equation}\label{eql16}
  \Delta {P}_{l} =
    P_i\otimes {\mathbbm 1} + {\mathbbm 1} \otimes P_i
\mathop{\longrightarrow}\limits_{ \kappa < \infty } \
   \ P_i \otimes e^{- \frac{P_0}{\kappa}} + {\mathbbm 1} \otimes P_i
   \end{equation}
   with the classical coproduct $\Delta P_0 = P_0 \otimes {\mathbbm 1} + {\mathbbm 1}
     \otimes P_0$ unchanged.
   The only cross-product relations which are modified look as follows
    $(M_{\mu\nu} = (M_i = \frac{1}{2} \epsilon_{ijk} M_{jk}, N_i = M_{i0}))$

\begin{equation}\label{eql17}
[N_i, P_j] = i \, \delta_{ij} P_0
\mathop{\longrightarrow}\limits_{ \kappa < \infty } \
  [N_i, P_j] = i \delta_{ij}
  [\frac{\kappa}{2} (1 - e^{-\frac{2P_0}{\kappa}})
  +\frac{1}{2\kappa}\overrightarrow{P}^{2}]
  +\frac{1}{\kappa}\, P_i P_j
\end{equation}
We introduce the following modified coproducts defining
$\kappa$-deformed cross-coproduct
\begin{eqnarray}\label{eql18}
 \Delta(N_i) =  N_i \otimes {\mathbbm 1} + e^{-\frac{P_0}{\kappa}} \otimes N_i
+\frac{1}{\kappa} \epsilon_{ijk} \, P_j \otimes M_\kappa \, .
\end{eqnarray}
The relations (\ref{eql16}-\ref{eql18}) are the only deformed
relations of $\kappa$-deformed D=4 Poincar\'{e} Hopf-Lie algebra.

 The D=4 superPoincar\'{e} algebra is obtained by adding two complex Weyl supercharges
 $Q_\alpha$ (chiral) and its Hermitean conjugates $\overline{Q}_{\dot{\alpha}}= Q_{\dot{\alpha}}$
  (antichiral), satisfying the relations
 \begin{eqnarray}\label{eql20}
\{Q_\alpha , \overline{Q}_{\dot{\beta}} \}  =  2 \{ \sigma^\mu P_\mu )_{\alpha \dot{\beta}}\, ,
\qquad
\{Q_\alpha , \overline{Q}_{{\beta}} \} =
\{ \overline{Q}_{\dot{\alpha}} , \overline{Q}_{\dot{\beta}} \}= 0
 \end{eqnarray}

  \begin{eqnarray}\label{eql21}
\{M_{\alpha \beta} , {Q}_{{\gamma}} \} & = &
\epsilon_{\alpha \gamma}  Q_\beta - \epsilon_{\beta \gamma} Q_\alpha
\qquad \{ M_{\alpha \beta}, \overline{Q}_{\dot{\gamma}} \} = 0
\\
\{M_{\dot{\alpha} \dot{\beta}} , \overline{{Q}}_{\dot{\gamma}} \} & = &
\epsilon_{\dot{\alpha} \dot{\gamma}}  \overline{Q}_{\dot{\beta}}
 - \epsilon_{\dot{\beta}\dot{\gamma}} \overline{Q}_{\dot{\alpha}}
\qquad \{ M_{\dot{\alpha} \dot{\beta}}, {Q}_{{\gamma}} \} = 0 \, ,
\\ \label{eql23}
[P_\mu, Q_\alpha ]& = & [P_\mu, \overline{Q}_{\dot{\alpha}} ] = 0
\end{eqnarray}

The D=4 Lorentz algebra can be described
 as $O(3,1) = Sl(2;c) \oplus \overline{Sl(2;c)}$ with the generators
$ M_{\alpha \beta} \sim M_i + i N_i \in Sl(2;c)$, $\overline{ M_{\dot{\alpha} \dot{\beta}}}
= M_i - i N_i \in \overline{Sl(2;c)}$.

Let us introduce the following graded bicrossproduct description of superPoincar\'{e}
algebra
\begin{equation}\label{eql24}
{\mathcal P}_{4;1} = (SL(2;c)\oplus \overline{SL(2;c)}
\ \ \raise.3ex\hbox{{\zul +} \kern-0.97em}\supset \
\overline{T}_{0;2}))\ltimes T_{4;2}
 \end{equation}
where $T_{0;2} = \overline{Q}_{\dot{\alpha}}$ and $T_{4;2} = (P_\mu , Q_\alpha)$ are graded
 Abelian sub-superalgebras.
The $\kappa$-deformation of the relation (\ref{eql24})
 leads to $\kappa$-deformed bicrossproduct structure \cite{kos}

\begin{equation}\label{eql25}
{\mathcal P}_{4;1}^{\kappa}
 = (SL(2;c)\oplus {SL(2;c)}
\ \  \raise.3ex\hbox{{\zul +} \kern-0.97em}\supset \
\overline{T}_{0;2}
\mathop{\triangleright\!\!\!\!\blacktriangleleft}\limits_{ \kappa } \
 T_{4;2}^{\kappa}
 \end{equation}
described as follows:

  i) We introduce the deformed coproducts for
  supercharges
  \begin{eqnarray}\label{eql26}
  \Delta^{(0)} Q_{\dot{\alpha}} = Q_\alpha \otimes {\mathbbm 1} +
 {\mathbbm 1}   \otimes Q_{\alpha}
\mathop{\longrightarrow}\limits_{ \kappa < \infty } \
\widetilde{\Delta} Q_{\alpha} = Q_{\alpha} \otimes e^{- \frac{P_0}{2\kappa}}
+ {\mathbbm 1} \otimes Q_{\alpha}\, .
  \end{eqnarray}
  The relation (\ref{eql26}) together with the formulae  (\ref{eql16}) and
  (\ref{eql23}) define the $\kappa$-deformed  graded-Abelian
   chiral Hopf superalgebra $T^{\kappa}_{4;2}$.

  ii) Besides the commutator (\ref{eql17}) the following cross-product relations of
  ${\mathcal P}_{4;1}$ in (\ref{eql24})  are $\kappa$-deformed

\begin{equation}\label{eql27a}
\{ Q_\alpha , Q_{\dot{\beta}} \} = 2(\sigma^\mu p_\mu )_{\alpha \dot{\beta}}
\mathop{\longrightarrow}\limits_{ \kappa < \infty } \
  \{ Q_\alpha , Q_{\dot{\beta}} \}
 = 4 \kappa \, \delta_{\alpha \dot{\beta}} \, sinh \frac{P_0}{2\kappa}
 - 2 e^{ \frac{P_0}{2\kappa}} (p_i \sigma_i)_{\alpha \dot{\beta}}
 \end{equation}

 \begin{equation}\label{eql27b}
 [N_i , Q_\alpha] = \frac{i}{2}(\sigma_i)_{\alpha}^{\ \beta} \, Q_{\beta}
\ \mathop{\longrightarrow}\limits_{ \kappa < \infty } \
  [N_i , Q_\alpha ]
 =
 \frac{i}{2} \, e^{-\frac{P_0}{\kappa}} (\sigma_i)_{\alpha}^{\ \beta}
 \, Q_\alpha +
 \frac{1}{2\kappa} \, \epsilon_{ijk} \, P_j (\sigma_k)_{\alpha}^{\ \beta}
 \, Q_\beta
\end{equation}

iii) The relations (\ref{eql18}), defining coalgebraic part of the
bicrossproduct are extended supersymmetrically as follows:
\begin{equation}\label{eql28}
\Delta N_i
\mathop{\rightsquigarrow}\limits_{ SUSY } \
\widetilde{\Delta} N_i = \Delta N_i - \frac{i}{4\kappa}
(\sigma_i)_{\alpha \dot{\beta}}
Q_{\alpha} \otimes  e^{\frac{P_0}{\kappa}} \, Q_{\dot{\beta}} \, .
\end{equation}
The formulae (\ref{eql18}) and (\ref{eql27a}-\ref{eql27b}) describe the deformed part
of the ${\mathcal P}_{4;1}^{\kappa}$ algebra, and the deformed coalgebraic part is
described by the relations (\ref{eql16}), (\ref{eql26}) and (\ref{eql28}).

Firstly standard $\kappa$-deformation of Poincar\'{e} algebra has been obtained by so-called
quantum AdS contraction of Drinfeld-Jimbo (DJ) $q$-deformation $U_q(O(3,2))$
of D=4 AdS Lie-Hopf algebra \cite{luk}.
We assume that in quantum AdS contraction procedure $q$ depends on $R$ in accordance
 with the asymptotic formula
 $ q(R)= 1 + ({\kappa \, R})^{-1} + {\mathcal O}( ({\kappa \, R})^{- 2})$,
 and
$ \lim\limits_{R\to \infty} \, U_{q(R)} (O(3,2)) = U_\kappa ({\mathcal P}_4)$
 provides the $\kappa$-deformed Poincar\'{e} algebra.
Similarly if we consider  $q$-deformed DJ superalgebra $U_q(OSp(1|4))$ one obtains
 \cite{luk2}

\begin{equation}\label{eql32}
\lim\limits_{R\to \infty} \, U_{q(R)} (OSp(1|4)) = U_\kappa ({\mathcal P}_{4;1}) \, .
\end{equation}

  In order to describe the deformation (\ref{eql32}) in the bicrossproduct basis
     one has to introduce the nonlinear change of basis.

It should be pointed out that recently there was obtained a two-parameter Jordanian
twist quantization $U_{\zeta_1, \zeta_2} (OSp(1|4))$ of D=4 AdS
superalgebra \cite{bor}. It
 was also shown \cite{bor}
  that by putting $\zeta_1 = \zeta_2 = \frac{1}{\kappa \, R}$
and performing the quantum contraction procedure
\begin{equation}\label{eq18bis}
\lim\limits_{R\to \infty} \, U_{\zeta_1, \zeta_2}  (OSp(1|4))\Big|_{\zeta_1 = \frac{1}{\kappa \, R}
\atop
\zeta_2 = \frac{1}{\kappa \, R}} = U^{LC}_\kappa ({\cal P}_{4;1})
\end{equation}
one gets the supersymmetric extension of so-called light-cone $\kappa$-deformaton
of Poincar\'{e} algebra (see e.g. \cite{bal,luk3})

\subsection{Final remarks}
It is known that new unified models of fundamental interactions are supersymmetric
(e.g. supergravities, superstrings, $M$-theory).
If the quantization of such theories generate the noncommutative structure of underlying
 spaces and superspaces, they also require quantum supersymmetries.
 In this note we presented more explicitely only one example of quantum
 supersymmetry, leading to Lie-algebraic structure of underlying
 noncommutative Minkowski spaces and superspaces which can be introduced
  as quantum representation spaces \cite{kos2}. Finally we add here that
   exists   also simpler way of
 deforming D=4 supersymmetries - by the twist function with a carrier spanned by
 the fourmomenta and supercharges (see e.g. \cite{kob}) - but it is
 outside of the scope of this short presentation.

\end{document}